\documentclass[letterpaper]{jpconf}
\usepackage{graphicx, amssymb}
\begin{document}
\title{Time-frequency analysis of extreme-mass-ratio inspiral signals in mock LISA data}

\author{Jonathan R Gair$^1$, Ilya Mandel$^{2,}$\footnote[3]
{Present address: Department of Physics and Astronomy, Northwestern University,
2131 Tech Dr.,  Evanston, IL} and 
Linqing Wen$^4$}

\address{$^1$ Institute of Astronomy, Madingley Road, CB3 0HA, Cambridge, UK}
\address{$^2$ Theoretical Astrophysics, California Institute of Technology, Pasadena, CA 91125}
\address{$^4$ Max-Planck-Institut f\"ur Gravitationsphysik, Am M\"uhlenberg 1, 14476,
Potsdam, Germany}

\ead{jgair@ast.cam.ac.uk, i-mandel@northwestern.edu, lwen@aei.mpg.de}

\begin{abstract}  
Extreme-mass-ratio inspirals (EMRIs) of compact objects with
mass $m \sim 1 - 10\ M_\odot$ into massive black holes with mass $M \sim 10^6\ 
M_\odot$ can serve as excellent probes of strong-field general relativity. The 
Laser Interferometer Space Antenna (LISA) is expected to detect gravitational
wave signals from $\sim 100$ EMRIs per year, but the data analysis of EMRI
signals poses a unique set of challenges due to their long duration and the
extensive parameter space of possible signals.  One possible approach is to
carry out a search for EMRI tracks in the time-frequency domain.  We have
applied a time-frequency search to the data from the Mock LISA Data Challenge (MLDC)
with promising results.  Our analysis used the Hierarchical Algorithm for
Clusters and Ridges to identify tracks in the time-frequency spectrogram
corresponding to EMRI sources.  We then estimated the EMRI source
parameters from these tracks.  In these proceedings, we discuss the results 
of this analysis of the MLDC round 1.3 data.
\end{abstract}

\section{Introduction}
The cores of most galaxies are expected to contain massive black holes (MBHs).  Compact stellar-mass objects (white dwarfs, neutron stars, and stellar-mass black holes) may be captured by an MBH, and may gradually spiral into the MBH under the influence of radiation reaction from the emission of gravitational waves (see \cite{Pau} and references therein for details).  Such extreme-mass-ratio inspirals (EMRIs) constitute one of the main sources for the planned gravitational-wave detector LISA (Laser Interferometer Space Antenna) \cite{LISA}.  Rate predictions are uncertain, but tens to thousands of EMRIs may be observed during the lifetime of the LISA mission \cite{Pau}.

Detection of EMRIs will not be an easy task, however.  The expected gravitational wave (GW) signals from EMRIs will be buried in instrumental noise and a foreground of galactic white-dwarf binaries.    Using the notation of \cite{AK}, EMRIs can be parametrized by: the two masses $M$ and $m$, the dimensionless spin parameter of the massive black hole $S$, the azimuthal frequency $\nu_0$ and orbital eccentricity $e_0$ specified at some moment of time, the orbital inclination $\lambda$, the source sky position angles $\phi_S$ and $\theta_S$, the orientation of the spin axis of the MBH $\phi_K$ and $\theta_K$, the distance to the source $D$, and three phase angles $\Phi_0$, $\gamma_0$ and $\alpha_0$, which specify, respectively, the initial azimuthal orbital phase, periapsis precession phase, and phase of the orbital-plane precession.  The large size of the EMRI parameter space, together with the long observation timescale ($\sim$ year), makes standard matched filtering impractical as a detection method.  

Time-frequency searches are one possible alternative to coherent matched filtering.  These have a much lower computational cost, albeit at the expense of lower detection sensitivity.  Time-frequency methods consist of building a spectrogram of the signal by dividing it into shorter segments and performing a Fourier transform on these segments, then identifying possible tracks in the spectrogram, and finally estimating source parameters from these tracks.  The pipeline we employ utilizes the Hierarchical Algorithm for Clusters and Ridges \cite{hacr07} to identify tracks by clustering bright pixels in binned spectrograms, followed by a simple parameter estimation routine that uses the track shape to infer six of fourteen EMRI parameters and the power variation over time to estimate the two sky position angles.

We have tested the performance of this time-frequency search on the Round 1.3 Mock LISA Data Challenge (MLDC) \cite{Arnaud1}. Challenges 1.3.1, 1.3.2, 1.3.3, 1.3.4, and 1.3.5 each included a single EMRI signal embedded in noise, with signal-to-noise ratio (SNR) between 40 and 110 \cite{Arnaud2}.  We attempted to analyze all five signals, and were successful in detecting EMRIs and estimating their parameters in the first four of these challenges.  We discuss our track search methods in section \ref{sec:search} and our parameter estimation in section \ref{sec:estimate}; the results are summarized in section \ref{sec:results}; and possible future improvements are discussed in section \ref{sec:future}.

\section{Track search methods}\label{sec:search}
The basis of our analysis of the data sets was the Hierarchical Algorithm for Clusters and Ridges (HACR). A detailed description of the method may be found in~\cite{hacr07}. HACR searches for clusters in the time-frequency spectrogram by identifying particularly bright ``black pixels'' (with power above the threshold $\rho_{\rm up}$) and then counting the number of relatively bright ``brown pixels'' (with power above the threshold $\rho_{\rm low}$) forming a contiguous cluster connected to each black pixel.  Any cluster containing more than $N_{\rm pix}$ pixels is a candidate event. The HACR search also employs binning --- we construct a sequence of binned spectrograms by using overlapping square bins of the form $2^{n_t} \times 2^{n_f}$ with $n_t, n_f = 0, 1, 2, ... $. This technique was first employed in the Excess Power search described in~\cite{wengair05,gairwen05,expowsymp}. We used the Excess Power search in parallel with HACR as a check of the results.

For the analysis of the MLDC Round 1.3 data, we tuned the thresholds of the HACR search for each challenge set using the procedure described in~\cite{hacr07}. This process involves injection of a test waveform into a sequence of noise realizations. For the test waveform, we used the corresponding MLDC training source in each case. The HACR thresholds were then determined by specifying an overall rate of false alarms per LISA mission (FAP) of 10\%. The training and challenge data sets (the LISA Simulator sets) were then analyzed as follows:
\begin{itemize}
\item The release data was used to construct $A$, $E$ and $T$ data streams.
\item The $A$, $E$ and $T$ data sets were bandpassed to a frequency range appropriate to that particular source, and then whitened using the theoretical noise spectral densities.
\item The data was divided up into segments of $2^{16}$ data points in length (approximately $11.4$ days). Each segment was FFT'd and thus a spectrogram of each data stream was constructed, with $64$ points in time and $32768$ points in frequency. For the 1.3.4 and 1.3.5 releases, segment lengths of $2^{14}$ and $2^{12}$ data points were also used.
\item The $A$ and $E$ spectrograms were summed pointwise and searched using the appropriate tuned HACR thresholds for an overall search false alarm probability of $10\%$.
\item The identified clusters were cleaned to aid parameter extraction. This cleaning used percolation (setting a high threshold and lowering it gradually to see features appear) to identify the brightest parts of the clusters, presumed to be from source tracks. This was followed by applying piecewise linear filters, aligned along these tracks, to remove spurious noise pixels from the cluster.
\end{itemize}
To test the HACR thresholds, we also constructed ``noise only'' data sets by subtracting the training signal from the MLDC training release and searching that data stream. In each case, the results were consistent with the imposed FAP. For challenges 1.3.1-1.3.4 we were able to identify several tracks in the challenge data set using HACR. In the challenge 1.3.5 data set, HACR identified a few pixels as interesting, but these could not be used for parameter extraction. This was consistent with the low stated prior on the source SNR $\in [40,60]$. We estimated the detection probability of the corresponding training data set at only $\sim85\%$.

\section{Parameter estimation methods}\label{sec:estimate}
The detection part of the search algorithm is fairly well developed. Parameter estimation from detected clusters and tracks is not so well advanced and is likely to improve significantly in the future. However, we describe the simple techniques that we have so far employed in this section. 

In the time-frequency domain, a typical EMRI signal is characterized by a sequence of tracks. These tracks correspond to harmonics of the azimuthal frequency, at frequencies $n\nu + \dot{\gamma}/\pi$ ($\nu$ is the Keplerian frequency and $\dot{\gamma}$ is the correction due to periapsis precession), plus sidebands that arise from orbital-plane precession which has characteristic frequency $\dot{\alpha}/(2\pi)$. The tracks appear in groups corresponding to a given value of $n$, since typically $\nu \gg \dot{\alpha}$. In general, we expect that the quadrupole ($n=2$) harmonics will be brightest for low-eccentricity inspirals. The separation between the groups of harmonics at a given time gives the value of $\nu$ at that time, while the separation between the members of a group gives an estimate of $\dot{\alpha}/(2\pi)$. The value of the frequency of a track, $f_{\rm t}$, then gives $\dot{\gamma}$, provided the harmonic is correctly identified (i.e., $\dot{\gamma}/\pi = f_{\rm t} - n\nu - k\dot{\alpha}/(2\pi)$, but $n$ and $k$ need to be known or guessed). The evolution of these frequencies depends on the source parameters. The track shape thus provides a constraint on the parameters. This idea was also discussed in~\cite{expowsymp}. The power fluctuation along a track arises from the motion of the LISA detector and thus constrains sky position.

\subsection{Sky position estimate}
The starting point for this estimate is the low frequency approximation to the LISA response given in~\cite{cutler98}. Making the simplifying assumption that the amplitudes of the two waveform polarizations are equal ($A_+  = A_\times$ in the notation of~\cite{cutler98}), the detector motion induced modulation, ${\cal M}(t)$, of the signal power, $h_I^2 + h_{II}^2$ can be approximated by 
\begin{equation}
{\cal M}(t) \propto (F_I^+)^2+(F_I^\times)^2+(F_{II}^+)^2+(F_{II}^\times)^2 \propto 1+ 6 \cos^{2} \theta + \cos^4\theta
\label{skymod}
\end{equation}
where
$$ \cos\theta = \frac{1}{2} \cos{\theta_S} - \frac{\sqrt{3}}{2} \sin{\theta_S}\cos\left(\frac{2\pi t}{T} - \phi_S\right)$$
in which our notation is consistent with~\cite{cutler98}, and $T = 1$ year.

To estimate the sky position, we took a $100\times 100$ grid of values of $(\theta_S, \phi_S) \in ([0,\pi/2], [0,2\pi])$. For each pair of values, we found via an automatic maximization the overall amplitude ${\cal A}$ that gave the best fit ${\cal A}{\cal M}(t)$ to the power profile along the track and computed the $\chi^2$ of the resulting fit. The best-fit sky position was the pair of parameters that minimized this $\chi^2$. Since Eq.~(\ref{skymod}) is based on a low frequency approximation, there is a degeneracy between the point ($\theta_S$, $\phi_S$) and the point ($\pi-\theta_S$, $\phi_S+\pi$), and so we always chose the value of $\theta_S$ that lay in the range $\left[0,\pi/2\right]$. We note also that the returned sky positions quoted in Table~\ref{restab} are $\pi/2 - \theta_S$ since $\theta_S$ as used by Cutler~\cite{cutler98} is the ecliptic colatitude not  ecliptic latitude.

\begin{figure}
\begin{center}
\vspace{-0.3in}
\includegraphics[keepaspectratio=true, width=4in, angle=-90]{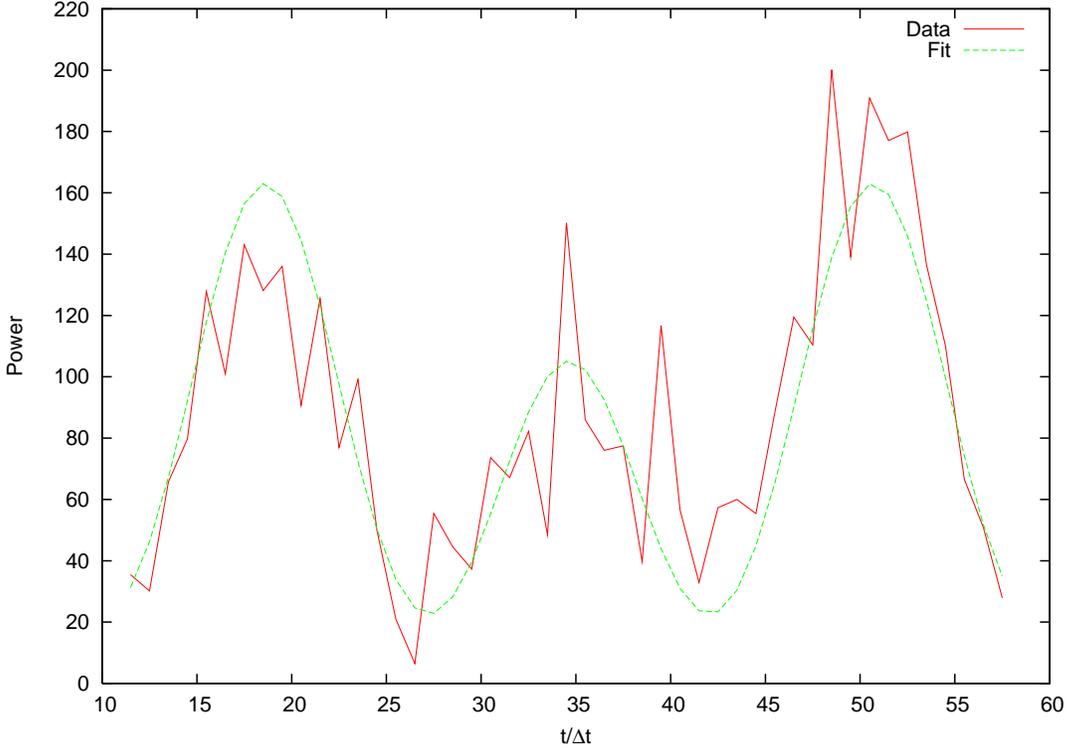}
\end{center}
\caption{\label{skyfig}The power profile along the longest detected track and the best-fit sky modulation curve for training data set 1.3.2.}
\end{figure}

For the Round 1.3 data sets, we used only the longest detected track for the sky position estimation. This could obviously be improved by using multiple tracks. For a single track, the sky position determination code runs in a fraction of a second. As an illustration of this method, we show in Figure~\ref{skyfig} the power profile along a track, and the best fit sky modulation curve for the longest track detected in an analysis of the 1.3.2 training data. The best fit parameters were $\theta_S = 1.78$ and $\phi_S = 3.64$. The actual sky position for that training set was $\theta_S = 1.83$, $\phi_S = 3.62$. In Table~\ref{restab} we also see that we recovered sky position well in the challenge sets, despite the approximations that went into Eq.~(\ref{skymod}).

\subsection{Other intrinsic parameters}
As mentioned above, the tracks give measurements of the three fundamental frequencies, $\nu$, $\dot{\gamma}$ and $\dot{\alpha}$. These depend on the six intrinsic parameters that are not phase angles --- $M$, $S$, $m$, $\lambda$,  $e_0$ and $\nu_0$. Explicit expressions for $\dot{\gamma}$ and $\dot{\alpha}$ in terms of these parameters are given in~\cite{AK}. Each measurement of a frequency gives a constraint on the six parameters but the number of independent measurements available from a set of tracks depends on the amount of frequency evolution of a source as well as on the length of track detected. To estimate parameters from the Round 1.3 data we attempted to find the 6 parameters that best matched the shape of the recovered tracks. We did not use the power in the tracks for anything other than sky position estimation. In principle the relative power in different $\nu$ harmonics is a measure of the eccentricity, while the power in different $\dot{\alpha}$ sidebands is a measure of the source orientation, $\theta_K$ and $\phi_K$. The total power of a track constrains $D$. In the future we plan to use the track power to go after $\theta_K$, $\phi_K$ and $D$. A time-frequency analysis will never have sensitivity to the remaining three phase parameters.

\section{Results}\label{sec:results}
We summarize the time-frequency properties that we determined for each of the challenge sets 1.3.1-1.3.4 in the following. Our recovered parameters and the true source parameters are compared in Table~\ref{restab}.

\paragraph{Challenge 1.3.1} We detected 5 frequency harmonics, two each for $n=2$ and $n=3$ plus one for $n=4$. The dominant harmonics for $n=2$ and $n=3$ were detected throughout most of the observation. The total frequency change was only $\sim 20$ frequency bins over the whole observation, which meant we could not determine the compact object mass to better than $5\%$. We were thus left with a degeneracy between $e$, $m$ and $M$. For the submitted results we fixed $m=9.5M_{\odot}$ and then used the detected tracks to determine the other parameters.

\paragraph{Challenge 1.3.2}Three harmonics were detected--- one each for $n=2,3,4$. Since we didn't measure any sidebands, we had no sensitivity to the value of $\dot{\alpha}$. The evolution equations for $\nu$ and $e$, and the value of $\dot{\gamma}/\pi$ depend only the quantity $S\cos\lambda$. We thus had some sensitivity to this combination, but not to $S$ and $\cos\lambda$ individually. Our results suggested $S\cos\lambda \approx 0$, and so for the submitted results we took $S=0.6$, i.e., the median of the allowed range, and $\lambda=\pi/2$.

\paragraph{Challenge 1.3.3}Tracks were identified from three different harmonics --- the dominant $n=2$ harmonic and a sideband, plus the dominant $n=3$ harmonic. However, none of these harmonics was detected for the whole inspiral --- the dominant $n=2$ harmonic was detected for $0 < t < 2.7 \times 10^7$s, the $n=2$ sideband was detected for $1.82\times 10^7$s $< t < 2.6\times 10^7$s and the $n=3$ harmonic for $1.62\times 10^7 $s $< t < 2.7\times 10^7$s.

\paragraph{Challenge 1.3.4}In this case, we did not detect the signal near $t=0$ nor near the plunge. We did detect the brightest (assumed to be $n=2$) harmonic, and one $\dot{\alpha}$ sideband for $2.15 \times10^7 $s $< t < 4.3\times 10^7$s, and we detected the $n=3$ harmonic plus one sideband for $2.15 \times10^7 $s $< t < 2.9 \times 10^7$s. Due to the low mass of the central black hole, these harmonics evolved significantly over the observed track, which provided more information for parameter extraction. The complication was in identifying which harmonics we were observing. We took the sidebands to be $k = 0, -1$ and then did a $\chi^2$ minimization over parameter space to find the sources that best reproduced the detected tracks.

\begin{table}
\begin{tabular}{|l|l|c|c|c|c|c|c|c|c|}
\hline \multicolumn{2}{|c|}{Challenge}&$\theta_{S}$&$\phi_S$&$S$&$m/M_{\odot}$&$M/M_{\odot}$&$\nu_0$ (mHz)&$e_0$&$\lambda$ \\\hline &Est.&0.25&0.31&0.68&9.5&9930000&0.1895&0.183&1.3 \\\cline{2-10} 1.3.1&True&0.301&0.318&0.664&9.602&9756157&0.1900&0.198&0.926 \\\cline{2-10} &Error&0.05&0.01&2.4\%&1.1\%&1.8\%&0.3\%&7.6\%&40\% \\\hline &Est.&0.5&0.22&0.6&9.54&5240000&0.32&0.253&1.57 \\\cline{2-10} 1.3.2&True&0.440&0.198&0.660&10.23&5194453&0.3192&0.262&1.266 \\\cline{2-10} &Error&0.06&0.02&9.1\%&6.7\%&0.9\%&0.3\%&3.4\%&24\% \\\hline &Est.&0.42&1.3&0.53&10.4&5070000&0.375&0.164&1.6 \\\cline{2-10} 1.3.3&True&0.316&1.263&0.548&10.42&4872374&0.3660&0.209&2.116 \\\cline{2-10} &Error&0.10&0.04&3.3\%&0.2\%&4.1\%&2.5\%&21\%&26\% \\\hline &Est.&0.2&5.3&0.58&10.02&952000&0.842&0.43&1.91 \\\cline{2-10} 1.3.4&True&-0.230&5.040&0.584&10.31&1021272&0.8426&0.402&1.865 \\\cline{2-10} &Error&0.43&0.26&0.7\%&2.8\%&6.8\%&0.1\%&7.0\%&2\% \\\hline\end{tabular}
\caption{Comparison of recovered parameters with true source parameters. Rows containing our recovered parameters are labeled by ``Est.'', rows of true parameters are labeled ``True'' and ``Error'' rows give the absolute difference for the angles $\theta_S$ and $\phi_S$ and the fractional difference $|\lambda_{\rm Est}/{\lambda_{\rm True}}-1|$ in percent for the other parameters.}
\label{restab}
\end{table}

\section{Summary and future plans}\label{sec:future}
The time-frequency algorithm performed well in the Round 1.3 analysis. The parameter recovery was good, although the exact parameter values are in some ways a poor measure of the algorithm performance. The method is better suited to determining combinations of parameters by measuring frequencies at certain times. To determine parameters, often several good constraints had to be combined with a poor one, leading to the appearance of errors in all the parameters. The errors in our parameter extraction thus hide the fact that we could measure combinations of parameters to high accuracy, which would be a useful input for follow up search algorithms. In the future, we hope to improve the technique in several ways.

\subsection{Track search improvements}
\begin{itemize}
\item The filtering algorithm described above will be formalized and automated.
\item We will use only the smaller bin sizes that are suitable for parameter extraction, rather then all possible bin sizes of the form $2^{n_t} \times 2^{n_f}$. 
\item Instead of binning, we will try running the search on multiple spectrograms constructed using different time segment lengths. Increasing the segment length increases the frequency resolution of the map, although this comes at a price: the source may evolve through several frequency bins in a single segment. The optimal bin sizes and segment length will therefore be parameter dependent.
\item The spectrogram construction will be improved. If the sky position is known, a sky-position optimized spectrogram can be constructed that gives more weight to $A$ or $E$ when the detector is favorably oriented for that sky location. Since we can estimate sky position quite well, it should be possible to do this construction.
\item Other possible improvements include using a higher specified FAP for the search and then rejecting noise clusters in post-processing, and binning using ``linear chirp'' bins rather than rectangular bins.
\end{itemize}

Eventually, we must also prepare for more realistic data sets containing multiple EMRI tracks that might cross each other.  In that case, our simple cluster cleaning technique will not work, and more sophisticated approaches will be needed to decide which of the multiple clusters should be associated with the same track.

\subsection{Parameter estimation improvements}
\begin{itemize}
\item Track power, not just track shape, will be utilized for parameter estimation. 
\item Markov Chain Monte Carlo methods will be used to find the parameters that best recover the detected time-frequency tracks.  This matched filtering of the recovered tracks will formalize the search method, and provide a way to quantify the errors in our parameter estimates.
\item We will follow up the t-f search with a matched filtering search to improve parameter estimates.
\end{itemize}

\ack
JG acknowledges support from St Catharine's College. IM would like to thank the Brinson Foundation, NASA grant NNG04GK98G and NSF grant PHY-0601459 for financial support. LW's work is supported 
by the Alexander von Humboldt Foundation's Sofja Kovalevskaja Programme funded by the German Federal Ministry of Education and Research.

\end{document}